\documentstyle[11pt,epsf]{l-aa}

\def\arcsec              {$^{\prime\prime}$} 
\def\arcs                {$^{\prime\prime}$}

\begin{document}
\thesaurus{11.17.3; 
	   11.17.4} 
\title{Sub-arcsecond imaging and spectroscopy of the radio-loud 
highly polarized quasar PKS~1610--771
\thanks{Based on observations obtained with the New Technology Telescope 
at the European Southern Observatory, La Silla, Chile}}
\author{F. Courbin\inst{1,2,3}, D. Hutsem\'ekers\inst{2\thanks{Also: 
chercheur  qualifi\'e    FNRS,   Belgium}},   G.    Meylan\inst{1}, P.
 Magain\inst{2\star\star}, S.G. Djorgovski\inst{4}}
\offprints{F. Courbin (Li\`ege address)}
\institute{
European Southern Observatory, 
Karl-Schwarzschild-Strasse 2, 
D--85748 Garching bei M\"unchen, Germany 
\and
Institut d'Astrophysique, Universit\'e de Li\`ege,
Avenue de Cointe 5, B--4000 Li\`ege, Belgium 
\and 
URA 173 CNRS-DAEC, Observatoire de Paris,
F--92195 Meudon Principal C\'edex, France
\and
Palomar Observatory, California Institute of Technology, 
Pasadena, CA 91125, USA\\}
\date{Submitted March 19$^{th}$; Accepted June 11$^{th}$}
\maketitle
\markboth{F. Courbin et al.: PKS~1610--771}{}
\begin{abstract}
We report on imaging and spectroscopic observations of the radio-loud,
highly polarized quasar PKS~1610--771   ($z$ = 1.71).   Our  long-slit
spectroscopy of the companion 4.55\arcs\ NW of the quasar confirms the
stellar nature of this object, so  ruling out the previously suspected
gravitationally lensed nature of this system.

PKS~1610--771 looks fuzzy on our  sub-arcsecond $R$ and $I$ images and
appears located in  a rich environment   of faint galaxies.   Possible
magnification, without image splitting  of the quasar itself,  by some
of these maybe foreground galaxies cannot be excluded.  The continuum
fuzz (made of the closest two objects, viz. A and D) is elongated in a
direction orthogonal to the $E$ vector of the optical polarization, as
in  high-redshift  radio-galaxies.   The   spectrum of   PKS~1610--771
appears strongly curved, in a convex way, with  a maximum of intensity
at $\sim$   7,600 \AA $    $ (2,800 \AA  $   $ rest  frame),  possibly
indicating a strong ultraviolet absorption by dust.
\end{abstract}
\section{Introduction}

In  the late eighties,  optical  imaging  surveys for  gravitationally
lensed quasars have used sample of known QSOs selected on the basis of
apparently large absolute luminosities and  high redshifts (Turner  et
al.  1984,  Meylan et al.   1990, Swings  et al. 1990,   Surdej et al.
1992).   Combining  all radio  and  optical  surveys,  the  number  of
multiple-image quasars due  to gravitational lensing amounts  to about
twenty more or less convincing cases (Keeton and Kochanek 1996).

In    the framework of the    ESO key-program on gravitational lensing
(Surdej et al.  1989), a  large sample of  highly luminous quasars has
been  observed  between 1989 and   1992.   One of  the gravitationally
lensed quasar candidates found is the radio-loud quasar PKS~1610--771,
at $z$ = 1.710 from Hunstead \&  Murdoch (1980), for which preliminary
poor-seeing  imaging  has been  obtained in  August  1991 by Meylan \&
Djorgovski using the   ESO 3.5-m New Technology   Telescope (NTT).  In
spite  of  these non-optimal seeing   conditions,  the object appeared
double.

In April and May 1995, we have been able  to obtain, under much better
seeing  conditions, new NTT   $R$ and  $I$   images  of the  field  of
PKS~1610--771, as well as long-slit low-resolution spectra of both the
quasar and   its  companion.   These  observations are   presented and
discussed here,  together with polarimetric  measurements  by Impey \&
Tapia (1988).

\section{Imaging}
\subsection{Observations and reductions}
The observations took place during  the two nights  of April 17-18 and
May 16-17, 1995, using the NTT  in remote control observing mode, from
the ESO Headquarters in   Garching, Germany.  The telescope   was used
with  EMMI, the ESO-Multi-Mode-Instrument.    The detector was the ESO
CCD \#36, a Tektroniks 2048$\times$2048, with a pixel size of 24$\mu$m
corresponding to 0.268\arcs\ on the plane of the sky.  The photometric
conditions were  good during  the  two observing runs, although  a few
cirri present during  the observations with  the  $I$ filter prevented
flux calibration of these frames.
\begin{figure}[t]
\begin{center}
\leavevmode
\epsfxsize=6.7cm
\epsffile{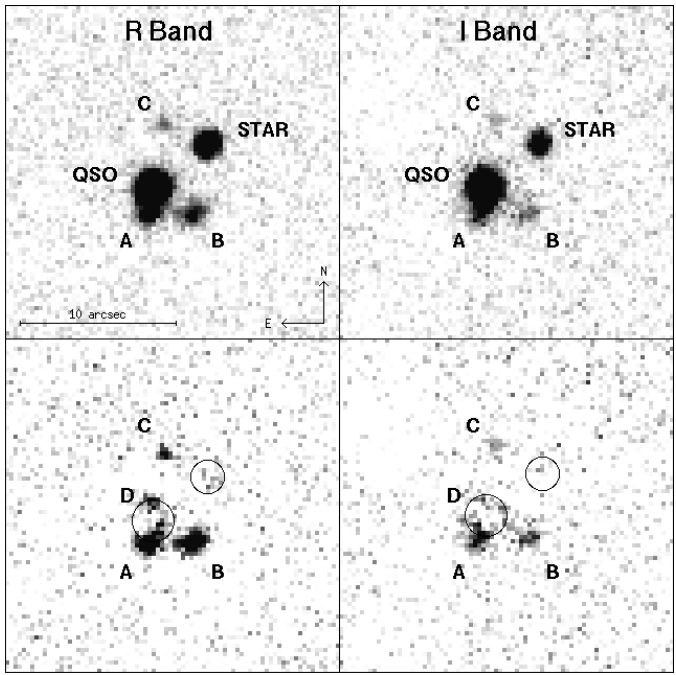}
\caption{Upper panels: NTT $R$ and $I$ band images of PKS~1610--771 
  obtained, in   both  filter, with  a total   exposure  time of 1,200
  seconds. The  seeing (FWHM) is 0.85\arcsec\  in $R$ and 0.95\arcsec\
  in $I$. Lower panels: same images
  as above,  but  a PSF has been   subtracted from the quasar  and its
  companion.   The two circles in both  PSF-subtracted images show the
  position of the quasar and the star.}
\end{center}
\end{figure}

Two images were obtained with the $R$ filter during the night of April
17-18, 1995.  The  final co-added frame has  a total exposure  time of
1,200  seconds, with  a  seeing of   0.85\arcsec,  giving a   limiting
magnitude of 23.5 in $R$ (2-$\sigma_{sky}$ in at least 3 pixels).

The two images in the $I$ filter were obtained during the night of May
16-17, 1995. The  final co-added frame has   a total exposure  time of
1,200 seconds, with a seeing of 0.95\arcsec.

Fig.~1 displays part of the images in the two $R$ and $I$ bands (upper
panels),  as well  as  the  result of  the   subtraction of  a  double
two-dimensional PSF profile (lower panels).  The  profiles of both the
quasar and the star  were fitted simultaneously  in order to take into
account relative light contamination of one  object by the other.  The
PSF subtraction was   carried  out using the   high-performance  codes
written by Remy (1996)  to detect gravitational deflectors.  In  order
to  optimize the quality   of the subtraction,  the numerical  PSF was
calculated by using four different stars, all of them as bright as the
quasar and as close to the quasar as possible in order to minimize any
possible PSF variations across the field.  The final PSF is a weighted
average of  the individual  PSFs  obtained from  these stars.  The PSF
subtractions  are carried out on the  individual 600 seconds exposures
in  order to avoid  any PSF changes due to  the alignment procedure of
the frames.  The individual residuals are then added and are displayed
in Fig.~1.\\
\subsection{Results}
In Fig.~1,   the  quasar   PKS~1610--771  appears  fuzzy  and  clearly
separated  from the   object  located at   4.55\arcs\ to  the   NW and
suspected to be a second, lensed, image of the quasar.  The latter has
definitely a stellar-like  profile and does  not leave any significant
residual after  the PSF   subtraction.   The follow-up   spectroscopic
observations (see  \S3 below) confirm its  stellar nature and rule out
the possibility that PKS~1610--771 is a multiple image gravitationally
lensed quasar.

Three resolved, galaxy-like, objects (noted A,  B, and C) surround the
quasar position on the  plane of the  sky. An additional  faint object
(noted D) appears only after  the PSF subtraction;  it is detected  in
the two  individual  $R$ frames and  their  sum,  as shown in  Fig.~1.
Although with a  worse seeing, significant  residuals are also seen on
the $I$ frames at the position where D is  detected on the $R$ frames.
This faint D feature   is unlikely to be an   artifact due to bad  PSF
subtraction, since no significant   residual  can be seen  after   the
subtraction   of the   PSF   profile from   the nearby  star.    Other
applications of our PSF subtraction method show it  to be reliable and
free of any artifacts (Magain et al. 1992, Remy et al.  1993).

Table~1 gives the $R$ magnitudes of the objects  in the field, as well
as their positions relative to the  quasar.  Flux calibration was done
using the star LTT~7987 (Landolt 1992) observed  at an airmass of 1.1,
while PKS~1610--771 was at a mean airmass of 1.55.  The magnitudes and
positions were obtained using  ``S-Extractor'', an aperture photometry
program  which computes  the  magnitudes through elliptical  isophotes
fitted to the  objects (Bertin \& Arnouts 1996;  Bertin 1996). Two $R$
magnitudes  are given: $m_{\it R}(1)$  magnitudes  are measured on the
summed $R$ frame while the $m_{\it R}(2)$  magnitudes are derived from
the  PSF subtracted frame.   The  magnitudes $m_{\it  R}(2)$ are  more
accurate for  the fuzzy objects, since they  do not suffer  from light
contamination by the point-like bright objects.

Although   we  cannot calibrate  the  $I$-band  image in  flux,  it is
possible to  derive the relative    brightness of the  quasar and  the
nearby  star, and to compare it  with the relative brightness found in
the  $R$   band.  We   find  $m_{I}(star)-m_{I}(quasar)=1.95$, whereas
$m_{R}(star)-m_{R}(quasar)=1.40$,  showing that  the   quasar is  much
redder than the star (as can, in fact,  be seen from the spectra shown
in Fig.~2).  We   cannot,   however, rule out    possible  photometric
variations of the quasar relative to the star, between April and May.

\begin{table}[t]
\begin{center}
\caption{Relative  positions  of the  five objects around  PKS 1610--771
along with their $R$ magnitudes (see text).  }
\vspace*{2mm}
\begin{tabular}{l l l l l}
\hline \hline
Object & $\Delta x$ & $\Delta y$ & $m_{\rm R}(1)$ & $m_{\rm R}(2)$\\ &
(\arcsec) & (\arcsec) & & \\
\hline  
Quasar & 0 & 0 & $18.2 \pm 0.20$ & ...\\ 
Star  & $+3.5 \pm 0.1$ & $+2.9 \pm 0.1$ & $19.6 \pm 0.20$ & ...\\ 
 A    & $-0.2 \pm 0.2$ & $-1.7 \pm 0.2$ & ... & $21.3 \pm 0.3$\\ 
 B    & $+2.6 \pm 0.2$ & $-1.6 \pm 0.2$ & $21.0 \pm 0.3$  & $21.3 \pm 0.3$\\
 C    & $+0.7 \pm 0.2$ & $+4.1 \pm 0.2$ & $23.0 \pm 0.5$ & $22.5 \pm 0.5$\\ 
 D    & $-0.1 \pm 0.2$ & $+1.0 \pm 0.2$ & ... & $23.0 \pm 0.5$\\
\hline
\end{tabular}
\end{center}
\end{table}
\section{Spectroscopy}

The spectra were obtained during  the same nights, using the long-slit
spectroscopic   capability  of EMMI.   The grism   \#1  (see EMMI User
Manual) provides low-resolution spectra  with a spectral resolution of
5.8 \AA\  pix$^{-1}$ and a scale   of 0.268\arcsec\  per pixel  in the
spatial direction.  
\subsection{The April 17-18, 1995 spectra}
The first aim of these observations being to investigate the potential
lensed nature of PKS~1610--771,  we acquired low resolution spectra of
the quasar and its stellar-like companion.  The 1\arcsec-wide slit had
a position angle  $\theta=320^{\circ}$ in order to obtain simultaneous
spectra of the two point-like objects.

Two 1,200-second exposures  and two 1,800-second exposures were taken,
under good seeing  conditions ($\sim$ 0.8\arcsec) at  airmass 1.5.   A
two-dimensional wavelength calibration     was  applied using    He+Ar
emission-line spectra.  The standard  star LTT~7987 was observed at an
airmass  of 1.1 and used to  calibrate the observations in flux (Stone
\& Baldwin 1983, Baldwin
\& Stone 1984, Landolt 1992).  Both the cosmic-ray removal and the sky
subtraction  were performed   on the   2-D  individual frames,  before
extracting   and   averaging  the   final  1-D    spectra,  which were
subsequently corrected from atmospheric extinction.

\subsubsection{Results}
The spectra of the  two objects are displayed  in Fig.~2. In the upper
panel is  the quasar spectrum showing a  very steep blue continuum, as
observed by Hunstead \& Murdoch (1980).  Longwards of $\sim$ 7,600 \AA
$ $ (2,800  \AA $  $  rest frame),  the  quasar spectrum decreases  in
intensity,  creating an unusual  convex shape.  We  identify the usual
MgII $\lambda  2,795$, CIII] $\lambda 1,909$,  and CIV $\lambda 1,549$
emission  lines, which yield the redshift  estimates given in Table 2,
in good agreement with previous results by Hunstead \& Murdoch (1980).

In  the lower  panel, the  spectrum  of the  stellar-like companion is
typical of a  late-type star with $T_{eff}\sim 4,500^{\circ}K$, ruling
out  the hypothesis of  gravitational splitting.  However, we can  not
exclude gravitational magnification  of the quasar's luminosity by the
nearby (angle-wise) fuzzy objects.

\begin{figure}[t]
\begin{center}
\leavevmode
\epsfxsize=9.0cm
\epsffile{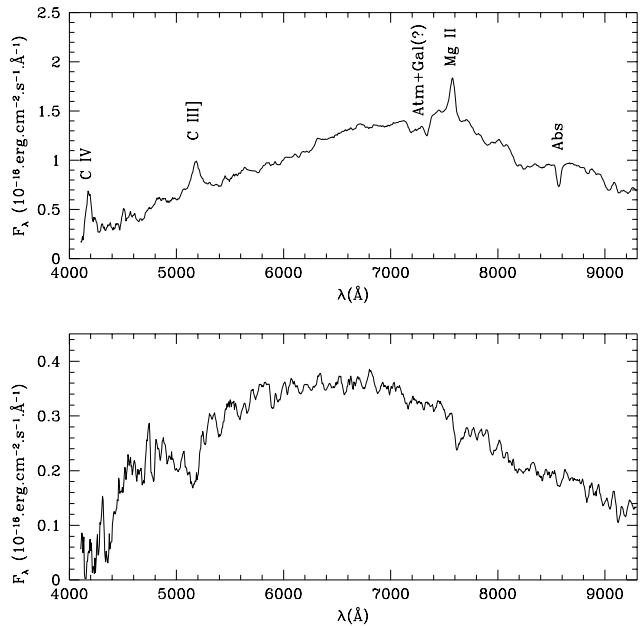}
\caption{April 1995 spectra of PKS~1610--771 (top) 
  and  of the nearby  stellar-like bright object (bottom).  The latter
  appears to be a  late-type star .   Note the general convex shape of
  the quasar continuum, and the steep  slope bluewards of $\sim$ 7,600
  \AA}
\end{center}
\end{figure} 
\subsection{The May 16-17, 1995 spectra}
We re-observed  PKS~1610--771,  in  May   1995,   this time  with    a
1.5\arcsec-wide slit positioned   along the  parallactic direction  in
order to avoid any possible bias caused by atmospheric dispersion.

We  also observed several standard  stars at different airmasses, some
close to the airmass of the quasar, in order  to have good corrections
for atmospheric absorptions.  The  flux calibration was done using the
standard star EG~274  (Stone \& Baldwin 1983,  Baldwin \& Stone  1984,
Landolt 1992) observed close to the quasar's airmass.

\begin{figure}[t]
\begin{center}
\leavevmode
\epsfxsize=9.0cm
\epsffile{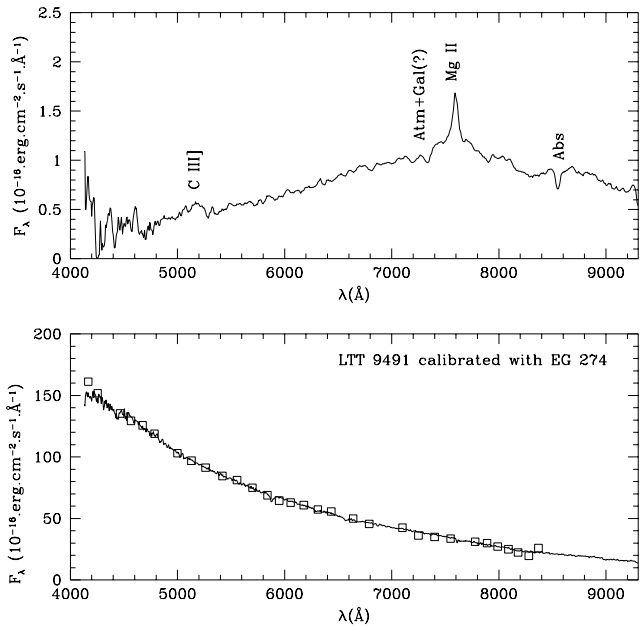}
\caption{
The May 1995 spectra  of PKS~1610--771 (top)  and of the standard star
LTT~9491    (bottom), both flux  calibrated    using the standard star
EG~274.   The tabulated values  of the absolute  flux  of LTT~9491 are
plotted with   open squares.   The    perfect agreement   between  the
reference and  observed  spectral  energy distributions   of  LTT~9491
show that     the unusual  convex  shape   of    the spectrum  of
PKS~1610--771  is intrinsic to  the quasar rather  than an artifact
due to data reduction.  }
\end{center}
\end{figure}
The  observation and data  reduction procedures  were similar to those
followed for the April 1995 data.  The  spectrum of PKS~1610--771 is a
stack of  four  1,800~second exposures (showing  all   the same convex
shape after flux calibration) and obtained  with seeing values between
1\arcsec\ and 1.2\arcsec.

The quasar spectrum is shown in Fig.~3 along with  the spectrum of the
standard star LTT~9491.  Both are flux  calibrated using the  standard
star EG~274.   The flux reference data for   EG~274, used to establish
the response curve of the  instrument, is  better sampled and  extends
over the  spectral range 4,000 - 12,000  \AA.  The  comparison between
the reference   and the observed    spectral  energy distributions  of
LTT~9491 illustrates a   good   accuracy  of our  flux     calibration
procedure.

\subsubsection{Results}

These May  1995  spectra of PKS~1610--771  confirm the  unusual convex
spectral  shape of  the continuum  observed in April  1995.  Since the
final reduced spectrum is better flux calibrated  than the April one -
but it has a  lower signal-to-noise ratio - we  measured the shape  of
the spectrum on the May data.  Note that there are obvious differences
in  the shape  of the  continuum    between April  and   May but  that
atmospheric refraction could  have affected the April spectra, because
of the slit orientation.  Assuming a power-law  dependence of the flux
density   with   $F_{\nu}\propto     \nu^{-\alpha}$,  we       measure
$\alpha_{b}\simeq 4.5$ from $\lambda$ 4,200 \AA $ $ to $\lambda$ 7,600
\AA, which is even  steeper than the  value $3.8$ reported by Hunstead
\& Murdoch (1980), and  $\alpha_{r} \simeq -1.0$ from  $\lambda$ 7,600
\AA $ $ to $\lambda$ 9,300 \AA.

An unidentified  absorption feature seen  in the April 1995 spectra is
detected at 8,552 \AA\ (and not present in the standard star, although
at the  same airmass as the  quasar).  Some additional absorptions are
possibly present between 7,200 and 7,400 \AA, on the  blue side of the
MgII  emission line (see Figs.~2  and  3).  These absorptions, blended
with  the atmospheric absorption  bands (also present  in the standard
stars  observed at the same  airmass as the  quasar),  are never fully
corrected  after flux calibration,  although the much  stronger A- and
B-bands  are perfectly removed.  This  suggests that these absorptions
are  actually present  in  the quasar   spectrum.  Note  that  all the
atmospheric absorption bands are  perfectly corrected in the  standard
star LTT~9491 (Fig.~3).
\begin{table}[t]
\begin{center}
\caption{Emission-line measurements and estimated redshift values from 
the spectra of PKS~1610--771} 
\vspace*{2mm}
\begin{tabular}{l|l l|l l}
\hline \hline
Line & \multicolumn{2}{c}{April 1995} & \multicolumn{2}{c}{May 1995} \\ &
$\lambda_{obs}$(\AA) & $z_{em}$ & $\lambda_{obs}$(\AA) & $z_{em}$ \\
\hline
CIV $\lambda 1549$ & 4184 & 1.701 & ... & ... \\ 
CIII] $\lambda 1909$ & 5185 & 1.716 & 5163 & 1.705 \\
MgII $\lambda 2795$ & 7573 & 1.706 & 7595 & 1.714 \\
\hline
\end{tabular}
\end{center}
\end{table}
\section{Discussion}

PKS~1610--771 was already recognized by Hunstead  \& Murdoch (1980) as
a flat-spectrum radio-source with a steep optical continuum.  Observed
with   the VLBI  by Preston et    al.  (1989), PKS~1610--771  appeared
slightly elongated (PA =  35 $^{\circ}$) at the  milliarcsecond scale.
Its high optical linear polarization (p = 3.8 $\pm$ 0.7 \%; $\theta$ =
78 $\pm$  5~$^{\circ}$) led Impey  \&  Tapia (1988,  1990) to classify
this object as   a  blazar,   although   there  is  no  evidence   for
polarimetric variability.

While  a few blazars display  a  curved continuum  (see  Falomo et al.
1994), none of  them  appears as steep  as  the one of  PKS~1610--771.
Baker  \& Hunstead  (1995)  give composite   spectra  of quasars  with
various values of $R$, the core-to-lobe radio flux density ratio.  The
composite        spectrum which corresponds     to the    $R$ ratio of
PKS~1610--771, viz.  $R$  $\sim$ 0.3  (Impey \&  Tapia 1990),  shows a
strong ultraviolet-optical bump.  Although the blue end of the bump is
well defined  (Baker \& Hunstead 1995), it  is not seen in our spectra
of PKS~1610--771, in which the slope of the short-wavelength continuum
(blueward 6,000 \AA $  $)  is constant  and much  steeper than in  the
composite   spectrum.   

Steep continua are not  uncommon in flat-spectrum radio quasars (e.g.,
Rieke et  al.    1979).   However,  such continua,   attributed to  an
extension of the  synchrotron radio emission,  extend from the optical
to the  infrared, contrary to  what is  observed   in the spectrum  of
PKS~1610--771. Indeed, the continuum slope we measure longwards of MgII,
$\alpha_{r}  \simeq -1.0$,   is  in   reasonable  agreement  with  the
radio-to-optical slope reported by Impey \& Tapia (1990): $\alpha_{ro}
\simeq   -0.9$, suggesting  a  net  deficiency  of   flux bluewards of
$\lambda_{rest} \sim 2,800$ \AA.  A most plausible explanation is dust
absorption, which apparently occurs  frequently in quasars (Webster et
al. 1995).  Dust may be  either internal, in  a dusty torus around the
nucleus or external,  i.e.   in the  host  galaxy, or  in  intervening
galaxies  on  the line of  sight.   Figure 4  illustrates a reasonable
example of de-reddening  using an SMC-like  extinction curve (Pr\'evot
et al.  1984) with E(B-V)=0.4 and assuming the dust at the redshift of
the quasar. The de-reddened spectrum shows a shape compatible with the
``usual'' shape of quasar continua, e.g. like  the quasars observed in
the LBQS sample (Morris et  al. 1991).  FeII emission   can also be  a
present, as observed for example in PG~0043+039 (Turnshek et al. 1994)

\begin{figure}[t]
\begin{center}
\leavevmode
\epsfxsize=9.0cm
\epsffile{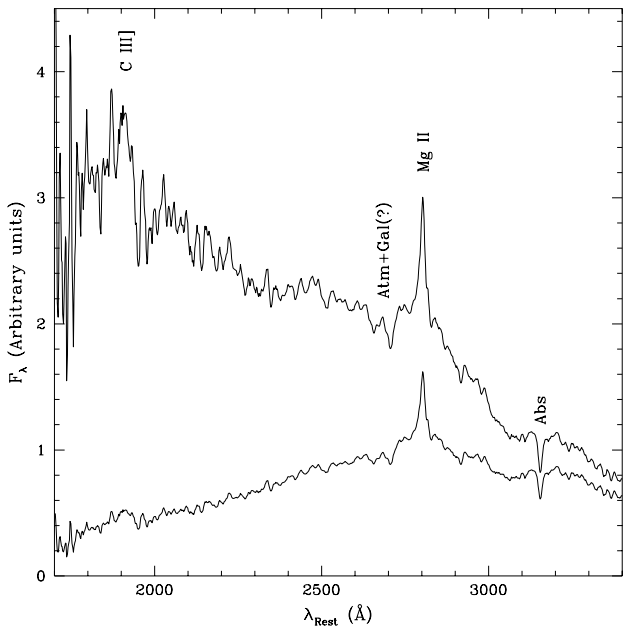}
\caption{The May 1995 spectrum of PKS~1610--771 (bottom) in the rest 
wavelength frame, and a plausible de-reddened version.}
\end{center}
\end{figure}
The  nature of  the fuzzy objects  (A+D) very  near   or superposed on
PKS~1610--771 is unclear.   They   could constitute the    quasar host
galaxy: their size  and brightness (cf.    Table 1) are comparable  to
those of the fuzzy objects  / host-galaxies found around high redshift
radio-loud quasars  (Hutchings 1995).  We  can however not exclude the
fact that it is a  galaxy located on  the  line of sight, although  no
absorption features  (in  the  MgII   line  for example) due  to    an
intervening  object  can be  unambiguously  identified in our spectra,
except the possible   residual absorptions observed  between 7,200 and
7,400 \AA.   Note however that these features  could as well be due to
the gap between  the two  strong  FeII  emission  bands seen in   AGNs
spectra at $\lambda_{rest}
\sim 2,000-2,600$~\AA $ $ and $\lambda_{rest}  \sim 2,700-3,000$~\AA $
$ (Wills et al.  1985). 
Even if we  can not ascertain  the redshift of the objects
around  PKS~1610--771, we  can give  the  plausible  redshift range of
1-1.7, comparing the magnitudes given  in Table  1 with the  predicted
ones from Guiderdoni and Rocca-Volmerange (1988).
  
Most interesting is the fact that the A-D fuzzy elongation is detected
in the continuum  and that its  direction (PA $\simeq$ 173$^{\circ}\pm
5^{\circ}$) is  orthogonal  to the $E$  vector of  the  --integrated--
optical  polarization  ($\theta$ $\simeq$  $78^{\circ}\pm  5^{\circ}$)
measured by  Impey   \& Tapia   (1988, 1990).   This  property  is  an
important characteristics  of high redshift  radio-galaxies (di Serego
Alighieri  et  al.  1993), which,   in the context  of AGN unification
models, are thought to be  related to radio-loud quasars, simply being
differently oriented re\-la\-tive to the observer.  This suggests that
the high polarization of PKS~1610--771 could be due to diffusion as in
radio-galaxies, rather than to  a synchrotron emission as in bona-fide
blazars.  Note that  the VLBI structure  of PKS~1610--771 (PA $\simeq$
35$^{o}$)  is  not  aligned  with the  optical    polarization, on the
contrary to the majority of  highly polarized radio-sources (Impey  et
al. 1991)

\section{Conclusions}
The present spectroscopic observations lead to the conclusion that the
quasar  PKS~1610--771  is   not gravitationally  lensed.   However, in
addition   to its  quite   unusual spectrum,   it  appears   fuzzy and
surrounded by  a   rich environment of  faint   galaxies  (which could
magnify the   quasar's luminosity).

The quasar's spectrum shows  a break at 7,600~\AA $  $ (2,800~\AA  $ $
rest frame) and an unusually steep slope  blueward 7,600~\AA $ $, most
likely due to  strong reddening by  intervening objects on the line of
sight and/or by the quasar's host galaxy.

Our sub-arcsecond    $R$ and $I$   images  of the quasar    show fuzzy
extensions supporting the possible   presence of absorbing  objects on
the line-of-sight.  The  orientation of the  continuum emission of the
fuzz around PKS~1610--771, perpendicular  to the optical polarization,
suggests  that the   polarization of  the    quasar could  be due   to
diffusion, as in radio-galaxies.

PKS~1610--771  therefore appears  as  an interesting  object, possibly
intermediate between blazars  and  radio-galaxies, its core  being not
completely  obscured  and hidden,  as for  example   in 3C~265 (Dey \&
Spinrad 1996), but highly reddened.

Further observations are  needed to identify the  exact nature and the
redshifts  of the fuzzy   objects around PKS~1610--771, its spectral
energy distribution on a  wide wavelength range, and more particularly
which mechanism is responsible for the observed optical polarization.

Confirming  objects  A+D  as  the host  galaxy  of PKS~1610--771  would
provide us with a unique case of highly  reddened (but not hidden) QSO
in a high redshift (z=1.7) radio  galaxy, strongly supporting the AGNs
unification schemes.

\begin{acknowledgements}
%
The  authors would  like to thank   R.W.  Hunstead and  J. Baker  for
helpful  comments and for  providing  their published composite quasar
spectrum, as  well as E.  Bertin  for help and comments concerning the
optimal use of his photometry program.
The SIMBAD   database  has been consulted  for   finding the available
literature on the object.
The research of the Li\`ege team on  gravitational lenses is supported
in part  by   the  contracts ARC   94/99-178   ``Action  de  Recherche
Concert\'ee de la    Communaut\'e Fran\c{c}aise'' (Belgium)   and  the
European Community, HCM Network CHRX-CT92-0044 .
SGD acknowledges a partial support from the NSF PYI award AST-9157412.\\ 
\end{acknowledgements}

\end{document}